\documentclass[aps,prb,twocolumn,longbibliography,superscriptaddress]{revtex4-1}
\usepackage{graphicx,amssymb,bm,amsfonts,amsmath}
\usepackage{natbib}

\begin{document}
\title{Role of polar compensation in interfacial ferromagnetism of LaNiO$_3$/CaMnO$_3$ superlattices}

\author{C. L. Flint}
\affiliation{Department of Materials Science and Engineering, Stanford University, Stanford, CA 94305, USA}
\affiliation{Geballe Laboratory for Advanced Materials, Stanford University, Stanford, CA 94305, USA}
\author{H. Jang}
\author{J.-S. Lee}
\affiliation{Stanford Synchrotron Radiation Lightsource, SLAC National Accelerator Laboratory, Menlo Park, CA 94025, USA}
\author{A. T. N'Diaye}
\author{P. Shafer}
\author{E. Arenholz}
\affiliation{Advanced Light Source, Lawrence Berkeley National Laboratory, Berkeley, CA 94720, USA}
\author{Y. Suzuki}
\affiliation{Geballe Laboratory for Advanced Materials, Stanford University, Stanford, CA 94305, USA}
\affiliation{Department of Applied Physics, Stanford University, Stanford, CA 94305, USA}

\begin{abstract}
Polar compensation can play an important role in the determination of interfacial electronic and magnetic properties in oxide heterostructures. Using x-ray absorption spectroscopy, x-ray magnetic circular dichroism, bulk magnetometry, and transport measurements, we find that interfacial charge redistribution via polar compensation is essential for explaining the evolution of interfacial ferromagnetism in LaNiO$_3$/CaMnO$_3$ superlattices as a function of LaNiO$_3$ layer thickness. In insulating superlattices (4 unit cells or less of LaNiO$_3$), magnetism is dominated by Ni--Mn superexchange, while itinerant electron-based Mn--Mn double-exchange plays a role in thicker metallic superlattices. X-ray magnetic circular dichroism and resonant x-ray scattering show that Ni--Mn superexchange contributes to the magnetization even in metallic superlattices. This Ni--Mn superexchange interaction can be explained in terms of polar compensation at the LaNiO$_3$--CaMnO$_3$ interface. These results highlight the different mechanisms responsible for interfacial ferromagnetism and the importance of understanding compensation due to polar mismatch at oxide-based interfaces when engineering magnetic properties. 
\end{abstract}    

\maketitle

Polarity mismatch at the interface of dissimilar materials has provided both challenges and opportunities for the heteroepitaxial growth of materials ranging from compound semiconductors to, more recently, complex oxides. In complex oxides, the multi-valent nature of the transition metal ions introduces the potential for electronic and atomic reconstruction regardless of how atomically precise is the interface. This electronic reconstruction has given rise to emergent behavior at the interfaces---from metallicity and superconductivity to ferromagnetism \cite{Ohtomo2004,Li2011}. Of particular interest has been the family of perovskite transition metal oxides (ABO$_3$) composed of AO and BO$_2$ stacks along the (001) direction. By incorporating perovskite oxides with different A- and B-site cation valences, one can introduce an electrostatic potential in the system. Mechanisms to alleviate this polarity-induced potential build-up may give rise to unexpected magnetic and electronic properties. The prototypical example is the formation of a 2-dimensional electron gas at the interface of LaAlO$_3$ and SrTiO$_3$ \cite{Ohtomo2004,Li2011}. Many other systems exhibiting emergent electronic phenomena at interfaces have been studied extensively since the discovery of metallicity at the LaAlO$_3$/SrTiO$_3$ interface. 

There have been significantly fewer studies demonstrating emergent magnetic behavior at oxide interfaces. For example, ferromagnetism at (111) LaFeO$_3$/LaCrO$_3$ interfaces is explained in terms of a 180$^{\circ}$ superexchange interaction between 3d$^5$ Fe$^{3+}$ and 3d$^3$ Cr$^{3+}$ ions which, according to the Goodenough Kanemori rules, should be ferromagnetic \cite{Ueda1998}. In digital LaMnO$_3$/SrMnO$_3$ superlattices, charge transfer from LaMnO$_3$ to SrMnO$_3$ gives rise to ferromagnetic double-exchange interactions\cite{Koida2002,Smadici2007}. Polar compensation is thought to partly drive this charge transfer\cite{Nanda2009}. However, charge transfer is also observed in systems without polar compensation\cite{Hoffman2013,Takahashi2001}. In LaNiO$_3$/SrMnO$_3$ superlattices, May \textit{et al.} found that no ferromagnetism is induced regardless of whether the superlattices exhibited metallic or insulating behavior \cite{May2009}, thus indicating that there is not enough charge transfer into SrMnO$_3$ to give rise to ferromagnetism even though a polar discontinuity is present. From these studies, it is clear that emergent ferromagnetism at interfaces has been associated with superexchange interactions or charge transfer driven exchange interactions. However the role of polar compensation in driving charge transfer that may give rise to emergent ferromagnetism is not clear. 

LaNiO$_3$ (LNO) is a promising material to explore the role of delocalized electrons as well as polar compensation in driving charge transfer in oxide heterostructures as it exhibits a thickness dependent metal-insulator transition \cite{Scherwitzl2011,King2014}. When LNO is combined with CaMnO$_3$ (CMO), ferromagnetism emerges at the interface and has been largely attributed to charge transfer from the LNO to CMO layer. In metallic superlattices, charge transfer driven by the leakage of itinerant electrons dominates. In insulating superlattices, charge transfer driven by polar compensation becomes apparent. Since the (001) CaO and MnO$_2$ layers are charge neutral, while the LaO and NiO$_2$ layers of LNO are positively and negatively charged respectively, polarity mismatch at the LNO/CMO interface also drives an interfacial charge redistribution to reduce the build-up of electric potential and contributes to the magnetic response. This effect should be present regardless of whether LNO is metallic or insulating but has not been observed in previous CMO-based superlattices \cite{Grutter2013}. 

\begin{figure}[t]
	  \includegraphics{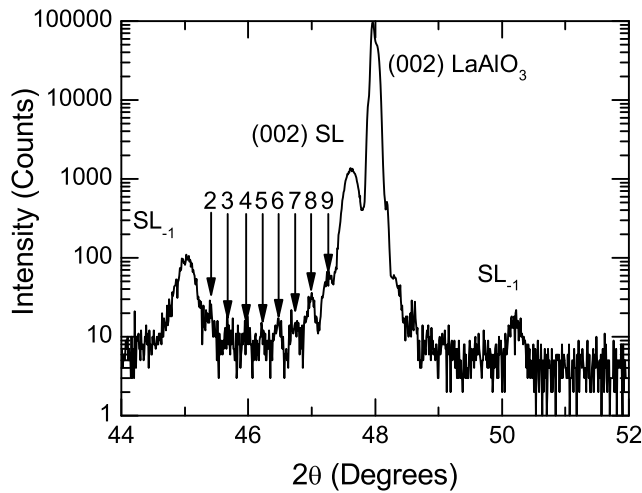}
\caption{2$\theta$-$\theta$ scan of an N=6, M=4 superlattice around the (002) LAO peak. Superlattice Bragg peaks and superlattice period thickness fringes are clearly seen, indicating high structural quality.}
    \label{fig:fig1}
\end{figure}

In this paper, we demonstrate that polar mismatch gives rise to interfacial charge redistribution and ferromagnetism in LNO/CMO superlattices. By focusing on LNO/CMO superlattices with only four unit cells of CMO, we are able to reduce the contribution of itinerant electron mediated ferromagnetic double exchange and highlight the polar compensation effect. In insulating superlattices, we have identified a small but significant ferromagnetic contribution from a Ni$^{2+}$--Mn$^{4+}$ superexchange interaction at the interface driven by polarity mismatch. In metallic superlattices, this contribution is combined with a ferromagnetic double exchange interaction that increases with LNO thickness. Together these results indicate that interfacial ferromagnetism is attributed to charge transfer driven by polarity mismatch as well as double exchange. 

To understand the role of polar mismatch in engineering interfacial ferromagnetic properties, we studied (LNO)$_N$/(CMO)$_M$ superlattices on 5 mm x 5 mm x 0.5 mm (001) LAO single crystal substrates, where N and M are the number of LNO unit cells and CMO unit cells per superlattice period, respectively. N was varied from 2 to 8, while M was held constant at 4. The superlattice periods were repeated 10 times. Films were deposited using a Coherent 248 nm KrF laser at 1 Hz with fluence of 1.3 J/cm$^{2}$. The background pressure was 60 mTorr of O$_2$ and the substrate was heated to 700 $^{\circ}$C. Unit cell growth was monitored \textit{in situ} via reflection high energy electron diffraction (RHEED). RHEED intensity oscillations were observed for each superlattice across all periods, indicating smooth layer-by-layer growth throughout every deposition. 

\begin{figure}[b]
	  \includegraphics{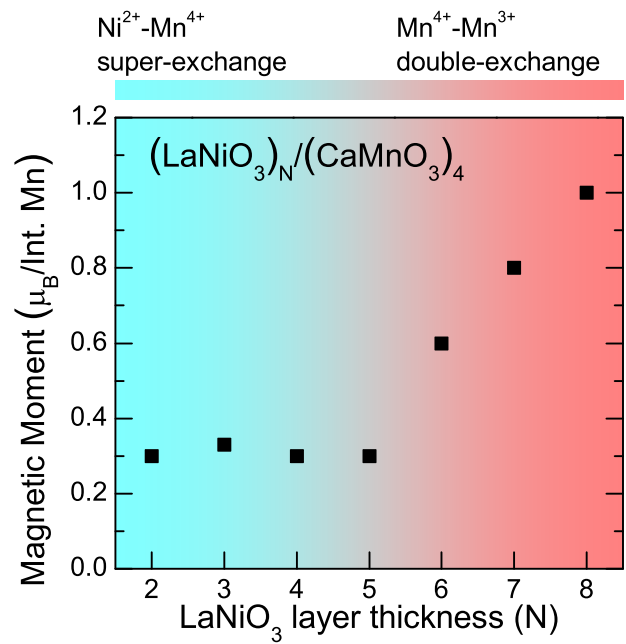}
\caption{(See online for color.) LNO layer thickness dependence of (LNO)$_N$/(CMO)$_4$ superlattice saturated magnetic moment at 2 T and 10 K. Note that even below the LNO metal--insulator transition a small ferromagnetic contribution remains. The background colors indicate the shift from primarily Ni$^{2+}$--Mn$^{4+}$ super-exchange at low N to Mn$^{3+}$--Mn$^{4+}$ double-exchange at high N.}
    \label{fig:fig2}
\end{figure}

All of our samples exhibited excellent crystallinity and layering as confirmed by x-ray diffraction (XRD). Figure 1 is a 2$\theta$-$\theta$ scan of an N=6, M=4 superlattice that exhibits clear superlattice Bragg peaks and superlattice period thickness fringes. Additionally, atomic force microscopy (AFM) of the superlattices revealed a surface roughness of less than half a unit cell, consistent with the  smooth growth of CMO at these conditions. Therefore RHEED, XRD, and AFM all confirmed high quality and precise control of the superlattice growth in this study.

Bulk magnetization measurements revealed ferromagnetic signal for all superlattices. Samples were field-cooled and measured at 10 K in fields up to 7 T in a Quantum Design Evercool Magnetic Properties Measurement System. Saturated magnetic moments for each superlattice are summarized in Fig. 2. Background subtraction was performed to isolate the film contribution to the magnetic signal from the LAO substrate contribution \cite{Khalid2010}. The magnetization has been normalized to the number of interfacial Mn ions. These results reveal that even superlattices with insulating layers of LNO exhibit ferromagnetism, nearly independent of LNO layer thickness at lower values of N.

Transport measurements of the superlattices indicated that superlattices with LNO layers of N$<$4 exhibited semiconducting or insulating behavior while superlattices with LNO layers of N$\geq$4 displayed metallic behavior. This behavior is consistent with previous studies of the thickness dependent metal--insulator transition in LNO thin films and superlattices \cite{May2009,Scherwitzl2011,Sakai2013}. Figure 3 shows resistivity versus temperature of a series of superlattices using a Quantum Design physical properties measurement system during cooling. As N increases, we observed a gradual approach to the bulk thin film LNO resistivity value, as we would expect for a system with finite thickness effects. For comparison, a 23 nm thick film of LNO is shown in Fig. 3. The resistivity of the 23 nm thin film is comparable to results of high quality LNO thin films reported elsewhere \cite{Sanchez2000,Scherwitzl2009,Son2010}. 

Let us first consider the saturated magnetic moment for metallic superlattices (N=4--8, M=4). The ferromagnetic moment in metallic superlattices primarily can be explained in terms of the leakage of electrons from the metallic LNO into the interfacial CMO layer in the form of a double-exchange interaction \cite{Nanda2007}. However the increasing saturated moment with increasing LNO layer thickness for metallic superlattices cannot be explained solely by the double-exchange interaction model which predicts a constant saturated moment as long as there is a metallic layer adjacent to the CMO layers. A plausible explanation could be that structural modification of the CMO crystal symmetry as a result of the increasing LNO layer thickness modifies the magnetism. For example, it has been known that in LNO/SMO superlattices as the ratio of LNO:SMO is increased, the superlattices take on more LNO-like bond angles \cite{May2011}. In the present case, the increased LNO:CMO ratio may lead to greater coherence across the LNO--CMO interface, leading to enhanced double-exchange ferromagnetism. Detailed structural characterization and correlation of the CMO crystal symmetry and the magnetism can be found elsewhere \footnote{C. L Flint, A. Vailionis, H. Zhou, H. Jang, J.-S. Lee, and Y. Suzuki (unpublished).}.

As the LNO thickness is decreased so that the superlattice is insulating, a double-exchange interaction among interfacial Mn ions due to leakage from itinerant electrons in the LNO cannot explain interfacial ferromagnetism. The ferromagnetism at N$<$4 must arise from another source due to the localized nature of electrons in insulating LNO. The presence of ferromagnetism in insulating superlattices seems at odds with previous work on M=8 LNO/CMO superlattices \cite{Grutter2013}. However a closer look at the transport properties of the superlattices from both studies indicates differences in residual resistivity, which may be correlated with oxygen stoichiometry and cation oxidation differences associated with the different oxygen growth conditions for the two studies. 

To identify the source of ferromagnetism in insulating superlattices, x-ray absorption spectroscopy (XAS) and x-ray magnetic circular dichroism (XMCD) in total electron yield mode (TEY) were performed at beamlines 4.0.2 and 6.3.1 at the Advanced Light Source of Lawrence Berkeley National Laboratory. L-edge XAS enables the determination of Mn and Ni valence and the corresponding XMCD allows for identification of magnetic elements. Samples were measured at an incident angle of 30$^{\circ}$ grazing in $\pm$1.5 T at 30 K. Figure 4(a)-(d) exhibit the Mn and Ni L-edge XAS and XMCD for N=2 and N=6 superlattices. The Ni and Mn XMCD in Fig.4b and d unequivocally demonstrate magnetic signal from both of these elements in insulating LNO/CMO superlattices. Similar results in both Ni and Mn XMCD spectra are found in metallic superlattices. The Mn XAS spectrum in Fig. 4(c) is consistent with Mn$^{4+}$ based on the L$_3$/L$_2$ intensity ratio and L$_3$--L$_2$ splitting \cite{Schmid2006}, as expected for CaMnO$_3$. The XMCD is consistent with Mn XMCD from La$_2$NiMnO$_6$, a double perovskite with Mn$^{4+}$ ferromagnetism \cite{Guo2009}. 

\begin{figure}[t]
	  \includegraphics{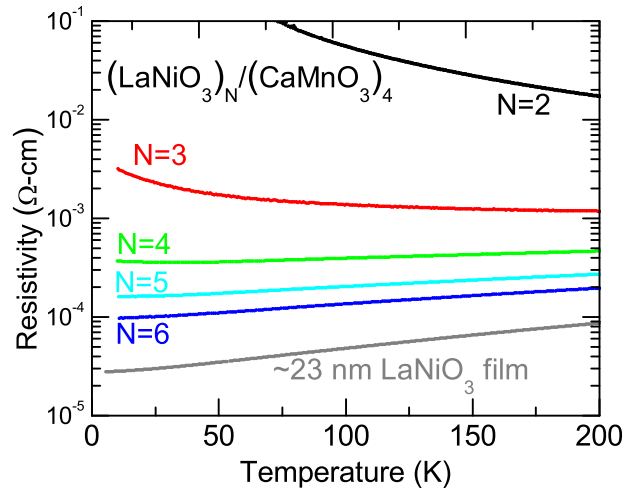}
\caption{(See online for color.) Temperature dependence from 10--200 K of superlattice resistivity for N=2--6 superlattices. Included is temperature dependence from 5--200 K of LNO thin film resistivity for comparison. Metal--insulator transition at N=4 and gradual approach to bulk LNO value is observed, consistent with previous results \cite{Scherwitzl2011,Grutter2013}.}
    \label{fig:fig3}
\end{figure}

\begin{figure*}[t]
  \includegraphics{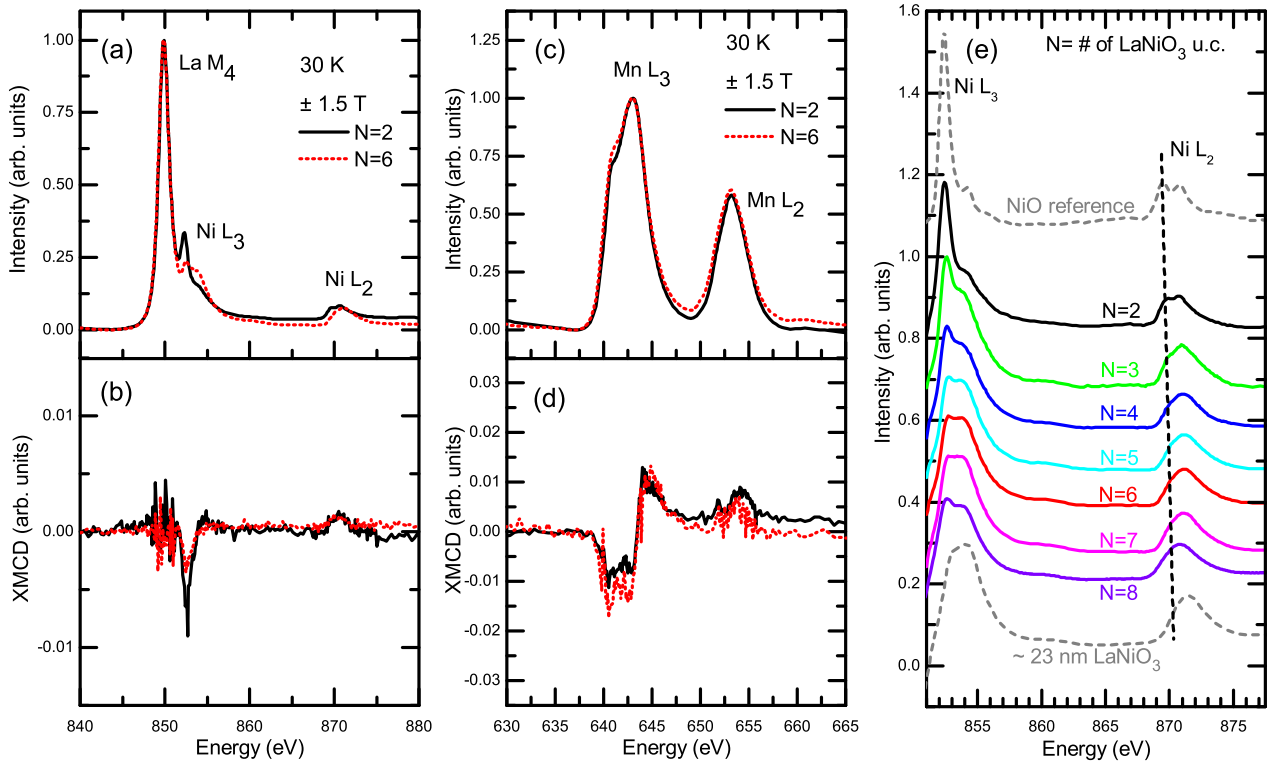}
 \caption{(See online for color.) (a) X-ray absorption spectrum of Ni L-edge at 30 K for an N=2, M=4 superlattice. (b) Corresponding Ni L-edge x-ray magnetic circular dichroism. (c) Mn L-edge x-ray absorption for the same superlattice. (d) Mn x-ray magnetic circular dichroism. (e) Ni L-edge X-ray absorption spectra of N=2--8 superlattices with La M$_4$ background subtracted (solid lines). Ni L-edge reference spectra of NiO (dashed grey line)(top) and an LNO thin film (dotted gray line)(bottom) for comparison. The shift in the L$_{2}$ edge is highlighted with the dotted line.}
 \label{fig:fig4}
\end{figure*}

A close look at the Ni L$_3$ edge reveals that it is obscured by the strong absorption intensity of the La M$_4$ peak and therefore care must be taken to make. In order to perform a direct comparison of the Ni L$_3$ edge peaks for varying LNO layer thickness, we normalized the La M$_4$ peak height to one and then fit to a combined Lorentzian and Gaussian expression, which was subsequently subtracted from the normalized data. The XAS results for the Ni L$_3$ edge (and the unaffected Ni L$_2$ edge) after subtraction can be found in Fig. 4(e), with offsets applied for clarity. Reference spectra for NiO and an LNO thin film are provided for comparison. The Ni L$_3$ edge peak after subtraction highlights a shift in valence across the metal--insulator transition. This shift can also be seen in the Ni L$_2$ edge, as indicated by the loss of the double peak feature and a gradual shift to high energy as emphasized by the dashed line. 

From comparison with the XAS spectra from NiO and LNO (Fig. 4(e), it is evident that the LNO in the thinnest superlattices has a significant fraction of Ni$^{2+}$, while the thickest metallic samples are nearly uniformly Ni$^{3+}$. Furthermore, the gradual shift from Ni$^{2+}$ to Ni$^{3+}$ as the LNO layer thickness increases suggests that the Ni$^{2+}$ is located at the interface since in thicker superlattices fewer interfaces are probed with XAS in surface sensitive TEY mode.  It is well known that Ni$^{2+}$--Mn$^{4+}$ 180$^{\circ}$ superexchange interactions are ferromagnetic based on the Goodenough--Kanamori rules and as observed in La$_2$NiMnO$_6$ \cite{Goodenough1958,Kanamori1959,Singh2009}. Therefore, the likely source of Ni--Mn ferromagnetism observed in these films is due to ferromagnetic superexchange interactions between the interfacial Ni$^{2+}$ and adjacent Mn$^{4+}$.

The XAS and XMCD results are consistent with interfacial Ni$^{2+}$ due to polar compensation arising from the differences in planar charge densities between CMO layers ([CaO]$^0$ and [MnO$_2$]$^0$) and LNO layers ([LaO]$^{+1}$ and [NiO$_2$]$^{-1}$). Recently, however, Johnston \textit{et al.} have proposed that Ni$^{2+}$ may be found in insulating nickelates due to oxygen--hole ordering \cite{Johnston2014}. In this scenario, negative charge transfer energy and electron-lattice coupling induce an ordered state in which half of the Ni is actually Ni$^{2+}$. In their model, the insulating state originates from coupling between rock--salt-like lattice distortions and ligand holes which allows for modulation of the Ni--O hybridization \cite{Johnston2014}. In addition to polar compensation effects, these aspects of nickelate physics must be considered. In our superlattices, then, the Ni$^{2+}$ fraction may be decreasing from thinner LNO layer superlattices to thicker LNO layer superlattices as a result of the thickness dependent metal--insulator transition and not due to polar compensation. The XAS trend would not be able to distinguish between these cases since it has limited structural sensitivity. 

To further investigate the presence of interfacial Ni$^{2+}$, we performed resonant x-ray scattering measurements (RSXR) at Beamline 13-3 at the Stanford Synchrotron Radiation Lightsource. RSXR has been used previously to explore charge transfer in manganite superlattices\cite{Smadici2007}. By performing scattering measurements using x-ray energies tuned to the Ni L$_2$ edge, we can get structural information with chemical and valence specificity. Furthermore, in superlattices with the proper symmetry, specific Bragg peaks can be probed to isolate interface contributions to the resonant x-ray scattering signal from those of the interior of the individual layers within the superlattice \cite{Park2013}. For example, in an M=4, N=4 superlattice, one would expect the (002) SL Bragg peak intensity to be reduced due to cancellation of the phases of the scattered wave. However, if there are differences between the interfaces and interior of an N=4, M=4 superlattice, the (002) peak would show enhanced interface resolution because only the difference between the bulk and interface would not cancel. For this purpose, an N=8, M=4 superlattice was studied because the symmetry is such that the (001) peak intensity is predominantly derived from the interior of the layers, while the (002) peak intensity is dominated by the interfaces\footnote{See Supplementary Material for details on RSXR analysis}. 

\begin{figure*}[t!]
  \includegraphics{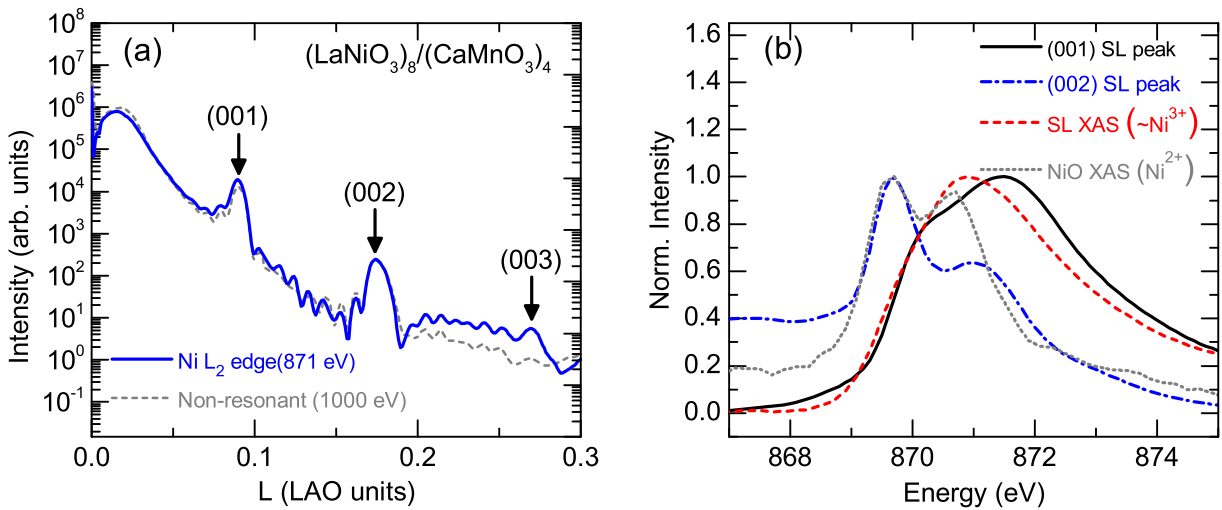}
 \caption{(See online for color.) (a) Specular X-ray reflectometry scan showing non-resonant (1000 eV) and Ni L$_2$ resonant (871 eV) reflectivity spectra of an N=8 superlattice. (b) Energy scans of (001) (solid black) and (002) (dash-dot blue) superlattice Bragg peaks. Simultaneously measured N=8 superlattice Ni L$_2$ XAS (dash red) and NiO Ni L$_2$ XAS (dot gray) are included for comparison.}\hfill
 \label{fig:fig5}
\end{figure*} 

A comparison of x-ray reflectivity measurements of an N=8 superlattice at non-resonant (1000 eV) and Ni L$_2$ resonant (871 eV) conditions provides insight into interface versus interior Ni valence (Fig. 5(a)). Superlattice Bragg peaks and thickness fringes are clearly observed. The (003) Bragg peak is absent due to symmetry of the superlattice.  By performing energy-dependent scattering measurements at the (001) and (002) Bragg peaks, we can observe Ni valence differences between the interior [(001) Bragg peak] and the interface [(002) Bragg peak], as seen in Fig. 5(b). Furthermore, by comparing scattering measurements at these resonant conditions to NiO and superlattice Ni XAS, we find that  the interior [(001)] of the LaNiO$_3$ layers is largely Ni$^{3+}$, while the interface [(002)] is predominantly Ni$^{2+}$(Fig. 5(b)). Together these results conclusively confirm the presence of interfacial Ni$^{2+}$, which is consistent with a Ni$^{2+}$--Mn$^{4+}$ superexchange ferromagnetism based on the Goodenough--Kanamori rules. 

One explanation for the origin of Ni$^{2+}$ in these superlattices, and one which is consistent with our experimental data, is the formation of Ni$^{2+}$ due to polar compensation between non-polar (001) CMO layers and polar (001) LNO. Such polar compensation has been observed previously in LNO/LAO superlattices and bilayers \cite{Liu2010,Middey2014}. In this scenario, the growth of LNO on CMO leads to a polar discontinuity and increasing electrostatic potential that drives the formation of interfacial oxygen vacancies. The existence of Ni$^{2+}$ induced by the metal--insulator transition, as proposed by Johnston \textit{et al.'s} \cite{Johnston2014}, may play some role in the insulating samples, but the persistence of the Ni magnetism and residual Ni$^{2+}$ in metallic samples is not fully consistent with this model. While the Ni$^{2+}$ represents a large portion of the Ni signal in thin, insulating films, it appears only to exist at the interfaces in the thickest, metallic films. 

In metallic samples, partial metallic screening might be expected to reduce the Ni$^{2+}$--Mn$^{4+}$ superexchange ferromagnetism contribution to the saturated magnetic moment by screening the build-up of a polarity-induced voltage. As a result, the contribution of itinerant electron-based double-exchange to the ferromagnetism becomes important.

Extrinsic oxygen off-stoichiometry -- to which perovskite oxides have a well-known susceptibility --  can also lead to Ni$^{2+}$. However, resonant x-ray scattering allows us to rule out uniformly oxygen deficient LNO as the origin of Ni$^{2+}$ due to the Ni$^{3+}$ signal from the interior of the LNO layer. Furthermore, the transport of superlattices with larger N reveal metallic behavior similar to that of stoichiometric LNO thin films.

Our experimental results support the model that Ni magnetism in these LNO/CMO superlattices is the result of Ni$^{2+}$--Mn$^{4+}$ superexchange ferromagnetism that results from interfacial charge redistribution due to polar compensation. Previous investigations into manganite-based interfacial ferromagnetism, utilizing both CRO and LNO as paramagnetic metals, have been explained in terms of itinerant electron-mediated double-exchange ferromagnetism. However, by reducing the contribution of the double-exchange interaction through a decrease in the CMO layer thickness, we have highlighted an additional important contribution to interfacial ferromagnetism due to polar compensation. This contribution does not exist in CRO/CMO or LMO/LNO superlattices, where both superlattice constituents have the same charge configuration. In LSMO/LNO and LCMO/LNO superlattices, this contribution would be obscured by charge transfer from the manganite and strong ferromagnetism of the manganite layer. Our studies indicate that the emergent ferromagnetic behavior at these interfaces is a delicate balance of superexchange and double exchange interactions. These interactions must be understood and taken into consideration as they will have important implications for future complex oxide heterostructure and device design by providing an additional path to engineering interfacial magnetism.

This research was funded by the Director, Office of Science, Office of Basic Energy Sciences of the U.S. Department of Energy under contract No. DE-SC0008505. Use of the Stanford Synchrotron Radiation Lightsource, SLAC National Accelerator Laboratory, is supported by the U.S. Department of Energy, Office of Science, Office of Basic Energy Sciences under Contract No. DE-AC02-76SF00515. 	The Advanced Light Source is supported by the Director, Office of Science, Office of Basic Energy Sciences, of the U.S. Department of Energy under Contract No. DE-AC02-05CH11231.

\bibliography{main}

\pagebreak
\clearpage
\onecolumngrid
\begin{center}
\textbf{\large Role of polar compensation in interfacial ferromagnetism of LaNiO$_3$/CaMnO$_3$ superlattices -- Supplementary Material}
\end{center}
\setcounter{equation}{0}
\setcounter{figure}{0}
\setcounter{table}{0}
\setcounter{page}{1}
\makeatletter
\renewcommand{\theequation}{S\arabic{equation}}
\renewcommand{\thefigure}{S\arabic{figure}}
\renewcommand{\bibnumfmt}[1]{[S#1]}
\renewcommand{\citenumfont}[1]{S#1}

\section{resonant x-ray scattering bragg peak analysis}

X-ray reflectivity from the superlattices can be modeled using kinematic formalism to determine the relative contributions of the "interior" and interfaces to the superlattice Bragg peak intensities. We can calculate these contributions beginning with the kinematic structure factor equation in 1-dimension:

\begin{equation}
F_{00L} = \sum_{i}^{N}f_{i}e^{i\cdot \pi \cdot L\cdot r_i}
\end{equation}

Since the growth order is first LNO and then CMO, the polar discontinuity occurs at that LNO/CMO interface (bottom/top). This is not important for the results obtained here, but it is a starting point for the indices. By separating the consituent unit cells into one-unit-cell interfacial segments, $f_{CMO_{int}}$ and $f_{LNO_{int}}$, as well as interior segments, $f_{CMO}$ and $f_{LNO}$, we can expand this equation as follows:

\begin{align}
F_{00L} &= f_{LNO}e^{i\cdot \pi \cdot L\cdot \frac{0}{12}} + f_{LNO}e^{i\cdot \pi \cdot L\cdot \frac{1}{12}} + f_{LNO}e^{i\cdot \pi \cdot L\cdot \frac{2}{12}} + f_{LNO}e^{i\cdot \pi \cdot L\cdot \frac{3}{12}} \nonumber \\ &\mathrel{\phantom{=}} + f_{LNO}e^{i\cdot \pi \cdot L\cdot \frac{4}{12}}+f_{LNO}e^{i\cdot \pi \cdot L\cdot \frac{5}{12}}+f_{LNO}e^{i\cdot \pi \cdot L\cdot \frac{6}{12}}+f_{LNO_{int}}e^{i\cdot \pi \cdot L\cdot \frac{7}{12}} \nonumber \\ &\mathrel{\phantom{=}} + f_{CMO_{int}}e^{i\cdot \pi \cdot L\cdot \frac{8}{12}}+f_{CMO}e^{i\cdot \pi \cdot L\cdot \frac{8}{12}}+f_{CMO}e^{i\cdot \pi \cdot L\cdot \frac{10}{12}}+f_{CMO}e^{i\cdot \pi \cdot L\cdot \frac{11}{12}}
\end{align}

Since we are using resonant x-rays, we can omit evaluation of the CMO structure factors. We can further group the $f_{LNO}$ and $f_{LNO_{int}}$ terms:

\begin{align}
F_{00L} &= f_{LNO_{int}}\left (e^{i\cdot \pi \cdot L\cdot \frac{7}{12}}\right ) \nonumber \\
&\mathrel{\phantom{=}}+f_{LNO}\left (e^{i\cdot \pi \cdot L\cdot \frac{0}{12}}+e^{i\cdot \pi \cdot L\cdot \frac{1}{12}}+e^{i\cdot \pi \cdot L\cdot \frac{2}{12}}+e^{i\cdot \pi \cdot L\cdot \frac{3}{12}}+e^{i\cdot \pi \cdot L\cdot \frac{4}{12}}+e^{i\cdot \pi \cdot L\cdot \frac{5}{12}}+e^{i\cdot \pi\cdot L\cdot \frac{6}{12}} \right ) 
\end{align}

Table 1 presents the evaluated scattering factor contribution to the structure factor for the (001), (002), and (003) Bragg peaks. The symmetric case of identical interfaces that differ from the interior is also included. The (001) Bragg peak is dominated by a bulk contribution. The (002) Bragg peak is solely interfacial contribution in the symmetric interfaces scenario, and is signficantly interfacial in nature in the asymmetric interfaces scenario. The relative amounts of interfacial and interior scattering factor contributions to the resonant soft x-ray scattering (RSXS) data in Fig. 5(b) explain the differences between the (001) and (002) Bragg peak RSXS, and confirm the interfacial origin of Ni$^{2+}$ in these superlattices.

\begin{table}[h]
\caption{Interface (Ni$^{2+}$) vs bulk (Ni$^{3+}$) contribution to resonant structure factor under the kinematic approximation under conditions of symmetric and asymmetric top and bottom LNO interfaces. }
\begin{ruledtabular}
\begin{tabular}{lll}
  Bragg Peak & Symmetric Interfaces & Asymmetric Interfaces \\ \hline
(001) & (0.13  $-$ 0.50i)*$f_{LNO_{int}}$ $-$
         (1.00  $-$ 3.73i)*$f_{LNO}$ & $-$(0.87  +  0.50i)*$f_{LNO_{int}}$ +
         (0.00  + 3.73i)*$f_{LNO}$ \\
   (002) & (1.50 + 0.87i)*$f_{LNO_{int}}$ +
         (0.00  + 0.00i)*$f_{LNO}$ & $\mathrel{\phantom{=}}$(0.50  + 0.87i)*$f_{LNO_{int}}$ +
         (1.00  + 0.00i)*$f_{LNO}$ \\
   (003) & (1.00  $-$ 1.00i)*$f_{LNO_{int}}$ $-$
         (1.00  $-$ 1.00i)*$f_{LNO}$ & $\mathrel{\phantom{=}}$(0.00  $-$ 1.00i)*$f_{LNO_{int}}$ +
         (0.00  + 1.00i)*$f_{LNO}$ \\
\end{tabular}
\end{ruledtabular}
\end{table}

\end{document}